\documentclass[aps,amsmath,preprint,epsf,superscriptaddress,nofootinbib]{revtex4-1}
\pdfoutput=1
\usepackage{graphicx,epsfig,amssymb,amsmath,color,hyperref,setspace,physics,cancel,feynmp}
\begin{document}
\draft \tighten
\title{To be, or not to be  finite?\\
The Higgs potential in Gauge-Higgs Unification}
\begin{flushright}
{\tt 
IPMU19-0104
}
\end{flushright}
\author{Junji Hisano}
\affiliation{Kobayashi-Maskawa Institute for the Origin
of Particles and the Universe, Nagoya University, Furo-cho Chikusa-ku, Nagoya, Aichi, 464-8602, Japan}
\affiliation{Department of Physics, Nagoya University, Furo-cho Chikusa-ku, Nagoya, Aichi, 464-8602, Japan}
\affiliation{Kavli IPMU (WPI), UTIAS, University of Tokyo, Kashiwa, Chiba 277-8584, Japan}
\author{Yutaro Shoji}
\affiliation{Kobayashi-Maskawa Institute for the Origin
of Particles and the Universe, Nagoya University, Furo-cho Chikusa-ku, Nagoya, Aichi, 464-8602, Japan}
\author{Atsuyuki Yamada}
\affiliation{Department of Physics, Nagoya University, Furo-cho Chikusa-ku, Nagoya, Aichi, 464-8602, Japan}

\begin{abstract}
In this paper, we investigate the finiteness of the Higgs effective
potential in an ${\rm SU}(\mathcal N)$ Gauge-Higgs Unification (GHU) model
defined on ${\bf M}^4\times S^1$. We obtain the Higgs effective
potential at the two-loop level and find that it is finite. We also
discuss that the Higgs effective potential is generically divergent for
three- or higher-loop levels. As an example, we consider an ${\rm SU}(\mathcal
N)$ gauge theory on ${\bf M}^5\times S^1$, where the one-loop
corrections to the four-Fermi operators are divergent. We find that the
Higgs effective potential depends on their counter terms at the
three-loop level.
\end{abstract}
\maketitle

\section{Introduction}
The Higgs mechanism is one of the essential ingredients in the standard
model (SM) of particle physics. It generates masses for the gauge bosons
and the fermions, which were forbidden by the gauge symmetries of the
standard model. Consequently, all the masses are described by the Higgs
vacuum expectation value (VEV) and the couplings, which is now in good
agreement with the Higgs coupling measurements at Large Hadron Collider
\cite{ATLAS:2018doi,CMS:2018lkl}.

In spite of the importance of the mechanism, the nature of the Higgs
boson has not been understood well. It has been discussed for a long
time that a scalar field is very sensitive to a UV cutoff scale, such as
the Planck scale or the grand unification scale, and it is not natural
that the Higgs VEV lies around the electroweak (EW) scale. If the Higgs
boson is really a fundamental scalar field, one needs to protect the
Higgs mass term from dangerous quantum corrections, which is greatly
achieved by supersymmetry
\cite{Veltman:1980mj,Davier:1979hr,Dimopoulos:1981au,Witten:1981nf,Dine:1981za}.
Alternatively, one can assume that the Higgs boson originates from
fields with non-zero spins. One such example is composite Higgs
models
\cite{Kaplan:1983fs,Kaplan:1983sm,Georgi:1984ef,Banks:1984gj,Georgi:1984af,Dugan:1984hq,Georgi:1985hf},
where the Higgs boson appears as a pseudo Nambu-Goldstone boson in
association with the condensation of fermions. Another example, which is
relevant to this paper, is the gauge-Higgs unification (GHU)
\cite{Fairlie:1979zy,Manton:1979kb,Forgacs:1979zs,
Hosotani:1983xw,Hosotani:1983vn,Hosotani:1988bm}, where the
four-dimensional gauge fields and the Higgs field are unified into gauge
fields in higher-dimensional spacetime.

In the GHU, we consider gauge theories defined on non-simply connected
spacetime and identify the Yang-Mills Aharonov-Bohm (AB) phases as Higgs
bosons.  Their tree level potential is protected because the Lagrangian
has to be invariant under gauge transformations. Since the
transformation variables need to be single-valued, not all of the gauge
transformations are consistent with the compactification of the
extra-dimensions. Thus, quantum corrections, which are sensitive to the
global structure of the spacetime, generate a Higgs potential that is
scaled by the compactification scale. It stabilizes the Higgs boson and
breaks the gauge symmetry dynamically. This is called the Hosotani
mechanism \cite{Hosotani:1983xw,Hosotani:1988bm}.

Although gauge theories are generically non-renormalizable in more than
four-dimensional spacetime, the Higgs potential might not depend on
UV-theory and might be completely determined within the framework of the
GHU, as conjectured in
\cite{vonGersdorff:2002rg,Hosotani:2005fk,Hosotani:2006nq}. In
fact, it has been explicitly shown that the Higgs potential is finite at
the one-loop level in generic GHU models
\cite{Hosotani:1983xw,Hosotani:1988bm,Davies:1988wt,Antoniadis:2001cv}
and at the two-loop level in an Abelian GHU model
\cite{Maru:2006wa,Hosotani:2007kn}. However, it has not been clear
whether the Higgs potential is finite at all orders.

To make things clear, let us state our criteria of finiteness. We allow
only the gauge interactions as irrelevant operators in the tree level
Lagrangian since they become relevant at the low energy four-dimensional
effective field theory. The counter terms for any other divergent
operators are assumed to be determined by UV-theory. From this
standpoint, we claim the Higgs potential is finite if all the
divergences are subtracted by the counter terms for the operators in the
tree level Lagrangian. In other words, the Higgs potential is claimed to
be divergent if it depends on any of the counter terms determined by
UV-theory.

In this paper, we investigate the finiteness of the Higgs potential
using an ${\rm SU}(\mathcal N)$ gauge theory defined on ${\bf M}^4\times
S^1$. Here, ${\bf M}^4$ represents the four-dimensional Minkowski
spacetime and $S^1$ represents a compactified extra-dimension. Although
it is the simplest manifold to realize the GHU, it is straightforward to
extend our discussion to other cases. 

To overcome technical difficulties that appear in perturbative
calculation, we discuss a method, {\it compactification by
superposition}, which greatly simplifies the calculation of the Higgs
potential in a non-Abelian GHU model. A similar method has been used in
the literature
\cite{Heffner:2015zna,Reinhardt:2016xci,Anber:2014sda,Ishikawa:2019tnw}
for Abelian cases. In this method, momentum sums and integrals in ${\bf
M}^4\times S^1$ are expressed as superposition of momentum integrals in
${\bf M}^5$, {\it i.e.} five-dimensional Minkowski spacetime. Thus, all
the AB phases can be ``gauged away'' from each integral. All the
information about the AB phases is then recovered when we superpose the
results after the integration. Another virtue of this method is that the
periodicity of the Higgs potential is manifest during the calculation,
which would become obscure if we adopted a straightforward calculation
with the Kaluza-Klein (KK) decomposition.

Using the method, we obtain the Higgs potential at the one-loop level
and that at the two-loop level, which turn out to be finite. We confirm
that the one-loop results agree with the previous works
\cite{Hosotani:1983xw,Hosotani:1988bm,Davies:1988wt} and the two-loop
results are consistent with those for an Abelian model
\cite{Maru:2006wa,Hosotani:2007kn}. The two-loop finiteness in a
non-Abelian model is highly non-trivial and is one of the new results in
this work.

To investigate the finiteness at higher-loop levels, we increase the
spacetime dimension and consider ${\bf M}^5\times S^1$, which allows
divergences to appear in an earlier stage of loop expansions. We find
that the four-Fermi operators are divergent at the one-loop level and
their counter terms contribute to the Higgs potential at the three-loop
level. Thus, the Higgs potential inevitably depends on UV-theory, which
falsifies the conjecture for this model.

This paper is organized as follows. In section \ref{sec_hosotani}, we
briefly review our setup and the Hosotani mechanism. In section
\ref{sec_cbys}, we explain our method to calculate the Higgs potential.
The one-loop and the two-loop calculations of the Higgs potential are
presented in section \ref{sec_one_two}. Then, we discuss the finiteness
of the Higgs potential at higher-loop orders in section
\ref{sec_higher}. Finally, we summarize in section \ref{sec_summary}.

\section{Dynamical Symmetry Breaking by Hosotani Mechanism}\label{sec_hosotani}
In this section, we review the Hosotani mechanism in an ${\rm SU}(\mathcal N)$
gauge theory defined on ${\bf M}^4\times S^1$. Here, ${\bf M}^4$ is the
four-dimensional Minkowski spacetime, whose coordinates are denoted by
$x^\mu$ with $\mu\in\{0,1,2,3\}$. The fifth dimension is compactified on
$S^1$ with radius $R$, whose coordinate is denoted by $y\in[0,2\pi
R)$. The gauge sector is described by a gauge coupling constant, $g$,
gauge bosons, $A_M^a$, and its field strength, $F_{MN}^a$, where the
capital indices, $M$ and $N$, run over $\{0,1,2,3,5\}$ and $a$ is the
group index. We also introduce massless Dirac fermions, $\psi_\ell$, in
arbitrary representations of ${\rm SU}(\mathcal N)$. In the Hosotani
mechanism, $A_5^a$ plays the role of the Higgs boson in the SM and its
VEV is denoted as
\begin{equation}
\langle A_5^a\rangle =\frac{\theta^a}{2\pi Rg},
\end{equation}
where $\theta^a$'s are constants.

In this paper, we use the background field methods \cite{Abbott:1980hw}
in order to evaluate the effective potential of $\theta^a$'s. For this
purpose, we separate $A_5^a$ into the quantum and background fields as
\begin{equation}
 A_5^a\rightarrow A_5^a+\frac{\theta^a}{2\pi Rg}.
\end{equation}

The Lagrangian we consider is given by
\begin{equation}
 \mathcal L =
 -\frac{1}{4} F_{MN}^aF^{aMN}
 +\sum_\ell\bar\psi_\ell i\gamma^MD_M\psi_\ell
 +\mathcal L_{\rm GF}+\mathcal L_{\rm ghost},
\end{equation}
where the gauge fixing terms are given by
\begin{equation}
 \mathcal L_{\rm GF}=-\frac{1}{2}\mathcal F^a\mathcal F^a,
\end{equation}
with
\begin{equation}
 \mathcal F^a=\partial^MA_M^a+\frac{f^{abc}}{2\pi R}A^b_5\theta^c.
\end{equation}
Here, $f^{abc}$ is the structure constant of ${\rm SU}(\mathcal N)$.
The corresponding Faddeev-Popov (FP) ghost terms are given by
\begin{equation}
 \mathcal L_{\rm ghost}=-\bar c^a
 \left[
 \partial^MD_M^{ab}
 -\frac{f^{ace}f^{bed}}{2\pi R}\theta^c\left(\frac{\theta^d}{2\pi R}+gA_5^d\right)
 \right]c^b.
\end{equation}
Here, the covariant derivative for an adjoint representation is given by
\begin{equation}
 D_Mc\equiv\left(\partial_M-ig A_M^aT^a-i\frac{\theta^aT^a}{2\pi R}\delta_M^5\right)c,
\end{equation}
where $[T^a]_{bc}=-if^{abc}$, and that for a fermion is given by
\begin{equation}
 D_M\psi_\ell\equiv
\left(
 \partial_M-ig A^a_M\tau_\ell^a
 -i\frac{\theta^a\tau_\ell^a}{2\pi R}\delta_M^5
 \right)\psi_\ell,
\end{equation}
where $\tau_\ell^a$ depends on the representation of $\psi_\ell$.

Throughout this paper, we adopt the following boundary conditions for
simplicity;
\begin{align}
 A_M^a(x^\mu,y+2\pi R)&=A_M^a(x^\mu,y),\\
 \psi_\ell(x^\mu,y+2\pi R)&=e^{i\beta_\ell}\psi_\ell(x^\mu,y),
\end{align}
where $\beta_\ell$'s are arbitrary phase factors.

Let us briefly explain the Hosotani mechanism using this setup.  Without the
boundary conditions, we could gauge away $\theta^a$'s by
\begin{align}
 A_5(x^\mu,y)&\to e^{-i\frac{\theta^aT^a}{2\pi R}y} A_5(x^\mu,y)e^{i\frac{\theta^aT^a}{2\pi R}y}
 -\frac{\theta^aT^a}{2\pi Rg},\label{eq_gaway1}\\
 \psi_\ell(x^\mu,y)&\to e^{-i\frac{\theta^a\tau_\ell^a}{2\pi R}y}\psi_\ell(x^\mu,y)\label{eq_gaway2},
\end{align}
where $A_M=A_M^aT^a$.  With the boundary conditions, however, we can
gauge away $\theta^a$'s only when
\begin{equation}
 e^{i\theta^aT^a}={\mathbb I},
\end{equation}
where $\mathbb I$ is the identity matrix. Due to this constraint, $\theta^a$'s
become physical degrees of freedom living in a compact space labeled by
$e^{i\theta^aT^a}$. Since the tree level Lagrangian is still invariant
under the transformation described by eqs.~\eqref{eq_gaway1} and
\eqref{eq_gaway2}, $\theta^a$'s do not have a potential at the tree
level. As we will see later, they obtain an effective potential at the
one-loop level and are stabilized. If some of $\theta^a$'s are non-zero
at the minimum of the effective potential, they dynamically break the
gauge symmetry and generate gauge boson masses.

\section{Compactification by Superposition}\label{sec_cbys}

In the usual computation of quantum corrections in a theory with
compactified extra-dimensions, we use the KK decomposition and evaluate
four-dimensional loop integrals for each KK mode. For example, in an
Abelian case, a typical integral at the one-loop level is given by
\begin{equation}
 I\equiv\frac{1}{2\pi R}\sum_{n=-\infty}^\infty\int \frac{d^4k}{(2\pi)^4}
 \left[
 k^\mu k_\mu-\left(\frac{n}{R}+\frac{\theta}{2\pi R}\right)^2
 \right]^{-s},\label{eq_typical}
\end{equation}
where $s$ is a positive constant and $n/R$ is the momentum along $S^1$,
which labels the KK modes.

Although the KK decomposition is useful in many cases, it is not in the
calculation of the effective potential of $\theta$, {\it i.e.} the Higgs
boson in the GHU. Since the Higgs boson is intrinsically the AB phase,
its effects can only be seen by particles that go around $S^1$ and
interfere with themselves. In the KK decomposition, however, it is
difficult to define the number of times the particles go around $S^1$
since the KK modes are momentum eigenstates.

In this paper, we discuss another way to decompose quantum fluctuations,
which has been used in
\cite{Heffner:2015zna,Reinhardt:2016xci,Anber:2014sda,Ishikawa:2019tnw}
for Abelian cases. The new decomposition is related to the KK decomposition by the
Poisson resummation formula\footnote{ It is essentially the same
transformation as is used in the previous calculations
\cite{Hosotani:1983xw,Hosotani:1988bm,Davies:1988wt}, where it is
applied after the four-dimensional integration. We apply it before the
four-dimensional integration and promote it to a five-dimensional one.
}, which is given by
\begin{equation}
 \sum_{n=-\infty}^{\infty}2\pi\delta\left(k_5-\frac{n}{R}\right)
 =2\pi R\sum_{m=-\infty}^{\infty}e^{-i2\pi Rmk_5}.\label{eq_conv}
\end{equation}
Using this identity, eq.~\eqref{eq_typical} becomes
\begin{equation}
 I=\sum_{m=-\infty}^\infty\int \frac{d^5k}{(2\pi)^5}e^{-i2\pi Rmk_5}
 \left[
 k^\mu k_\mu-\left(k_5+\frac{\theta}{2\pi R}\right)^2
 \right]^{-s}.
\end{equation}
It implies that a loop integral in ${\bf M}^4\times S^1$ can be
reproduced by superposition of loop integrals in ${\bf M}^5$. Since the
phase factor is the shift operator of $(x^\mu,y)\to (x^\mu,y-2\pi Rm)$,
we call $m$ the winding number.

Since the AB phase can be ``gauged away'' in ${\bf M}^5$, we can further
simplify the integral as
\begin{equation}
 I=\sum_{m=-\infty}^\infty e^{i\theta m}
 \int \frac{d^5k}{(2\pi)^5}e^{-i2\pi Rmk_5}
 \left[k^M k_M\right]^{-s},
\end{equation}
by shifting $k_5$. In this expression, all the $\theta$-dependence
appears as phase factors in association with the superposition and we can
execute the loop integrals independently of $\theta$. Furthermore, the
periodicity of $\theta$ is manifest.

The above decomposition is very powerful especially in a non-Abelian
case, where we have the following identity;
\begin{equation}
 \frac{1}{2\pi R}\sum_{n=-\infty}^{\infty}S
 \left(\frac{n}{R}+\frac{\Theta}{2\pi R}\right)
 =\sum_{m=-\infty}^{\infty}e^{i\Theta m}
 \int_{-\infty}^{\infty} \frac{dk_5}{2\pi}
 e^{-i2\pi Rk_5m}S(k_5),\label{eq_conv2}
\end{equation}
where $\Theta$ is a Hermitian matrix and $S(\dots)$ is an analytic function
or its extension to a matrix function (we call it as an ``analytic
function'' in short). We provide its proof in appendix
\ref{apx_proof_main}. It removes all the matrix-valued objects from
momentum integrals and simplifies the calculation enormously.

In this paper, we do not try to construct the Feynman rules that
generate the final expressions directly. Instead, we first use the KK
decomposition and then convert the expressions by
eq.~\eqref{eq_conv2}.

\section{Higgs Effective Potential up to Two-loop Level}\label{sec_one_two}
In this section, we calculate the one-loop and the two-loop effective
Higgs potentials explicitly and show that they are finite.
\subsection{One-loop Effective Potential}
At the one-loop level, the quantum corrections to the Higgs effective
potential from the gauge bosons, the FP ghosts and the fermions can be
calculated from
\begin{align}
 V^{\rm 1L}_{A,\rm eff}(\theta) =i
\begin{fmffile}{1L-A}
\begin{gathered}
\begin{fmfgraph}(30,30)
\fmfleft{i}
\fmfright{o}
\fmf{phantom,tension=5}{i,v1}
\fmf{phantom,tension=5}{v2,o}
\fmf{photon,left,tension=0.4}{v1,v2,v1}
\end{fmfgraph} 
\end{gathered}
\end{fmffile}
&=-\frac{5i}{2}\frac{1}{2\pi R}\sum_n\int\frac{d^4k}{(2\pi)^4}
 \tr\ln\left[
 k^2-\left(\frac{n}{R}+\frac{\theta^aT^a}{2\pi R}\right)^2
 \right],\nonumber\\
 V^{\rm 1L}_{c,\rm eff}(\theta) =i
\begin{fmffile}{1L-C}
\begin{gathered}
\begin{fmfgraph}(30,30)
\fmfleft{i}
\fmfright{o}
\fmf{phantom,tension=5}{i,v1}
\fmf{phantom,tension=5}{v2,o}
\fmf{ghost,left,tension=0.4}{v1,v2,v1}
\end{fmfgraph} 
\end{gathered}
\end{fmffile}&=i\frac{1}{2\pi R}\sum_n\int\frac{d^4k}{(2\pi)^4}\tr\ln
 \left[
 k^2-\left(\frac{n}{R}+\frac{\theta^aT^a}{2\pi R}\right)^2
 \right],\nonumber\\
 V^{\rm 1L}_{F,\rm eff}(\theta) =i
\begin{fmffile}{1L-F}
\begin{gathered}
\begin{fmfgraph}(30,30)
\fmfleft{i}
\fmfright{o}
\fmf{phantom,tension=5}{i,v1}
\fmf{phantom,tension=5}{v2,o}
\fmf{fermion,left,tension=0.4}{v1,v2,v1}
\end{fmfgraph} 
\end{gathered}
\end{fmffile}&=\sum_\ell2i\frac{1}{2\pi R}\sum_n\int\frac{d^4k}{(2\pi)^4}
 \tr\ln\left[
 k^2-\left(\frac{n}{R}+\frac{\theta^a\tau_\ell^a-\beta_\ell}{2\pi R}\right)^2
 \right],
\end{align}
respectively. Here after, all the sums except for those of the flavor
index, $\ell$, are taken from $-\infty$ to $\infty$ if not explicitly
specified. We convert them with eq.~\eqref{eq_conv2} as
\begin{align}
 &\frac{1}{2\pi R}\sum_n\int\frac{d^4k}{(2\pi)^4}\tr\ln
 \left[
 k^2-\left(\frac{n}{R}+\frac{\Theta}{2\pi R}\right)^2
 \right]\nonumber\\
 &\hspace{3ex}=-\lim_{s\to0}\frac{d}{ds}\frac{1}{2\pi R}\sum_n\int\frac{d^4k}{(2\pi)^4}\tr
 \left[
 k^2-\left(\frac{n}{R}+\frac{\Theta}{2\pi R}\right)^2
 \right]^{-s}\nonumber\\
 &\hspace{3ex}=-\lim_{s\to0}\frac{d}{ds}\sum_m
 \tr e^{i\Theta m}\int\frac{d^5k}{(2\pi)^5}e^{-i2\pi Rmk_5}(k^Mk_M)^{-s},
\end{align}
where $\Theta=\theta^aT^a$ or
$\Theta=\theta^a\tau_\ell^a-\beta_\ell$. The loop integrals are executed
in appendix \ref{sec_integral} and we get
\begin{align}
 \lim_{s\to0}\frac{d}{ds}\int\frac{d^5k}{(2\pi)^5}e^{-i2\pi Rmk_5}(k^Mk_M)^{-s}
 &=\frac{3i}{128|m|^5\pi^7R^5},
\end{align}
for $m\neq0$. Thus, we get
\begin{equation}
 V^{\rm 1L}_{\rm eff}(\theta)
 =-\frac{9}{256\pi^7R^5}\sum_{m\neq0}\frac{1}{|m|^5}\tr e^{i\theta^aT^am}
 +\frac{3}{64\pi^7R^5}\sum_\ell\sum_{m\neq0}\frac{1}{|m|^5}\tr e^{i(\theta^a\tau_\ell^a-\beta_\ell)m}+C,
\end{equation}
where $C$ represents the $\theta$-independent divergent terms, {\it
i.e.} the contributions from $m=0$. The $\theta$-dependent part is
finite and consistent with the previous works
\cite{Hosotani:1983xw,Hosotani:1988bm,Davies:1988wt}.

\subsection{Two-loop Effective Potential}
At the two-loop level, we need to work a little more because we cannot
directly use eq.~\eqref{eq_conv2} to convert expressions. In the
calculation, we often face the following expression;
\begin{equation}
 \frac{1}{(k+p)^\mu (k+p)_\mu-\left(\frac{n+n'}{R}+\frac{\theta^a\tau^a-\beta}{2\pi R}\right)^2}
 \tau^b\frac{1}{k^\mu k_\mu-\left(\frac{n}{R}+\frac{\theta^a\tau^a-\beta}{2\pi R}\right)^2},
\end{equation}
where $\tau^a$'s are generators of ${\rm SU}(\mathcal N)$.  It is not an analytic
function of $\left(\frac{n}{R}+\frac{\tau^a\theta^a-\beta}{2\pi
R}\right)$ since we have $\tau^b$ in the middle. To remove $\tau^b$, we
use
\begin{equation}
 S(\theta^a\tau^a)\tau^b=\tau^c\left[S\left(\theta^a\tau^a+\theta^aT^a\right)\right]_{cb},\label{eq_move}
\end{equation}
where $S(\dots)$ is an arbitrary analytic function. Here, the indices in
the subscript are those for $T^a$, not for $\tau^a$. Its proof is given
in appendix \ref{apx_proof_com}. Then, the expression becomes
\begin{equation}
 \tau^c\left[
 \frac{1}{(k+p)^\mu (k+p)_\mu-\left(\frac{n+n'}{R}+\frac{\theta^a\tau^a-\beta +\theta^aT^a}{2\pi R}\right)^2}
 \frac{1}{k^\mu k_\mu-\left(\frac{n}{R}+\frac{\theta^a\tau^a-\beta}{2\pi R}\right)^2}
 \right]_{cb}.
\end{equation}
Now, the inside of the square brackets can be seen as an analytic
function of $\left(\frac{n}{R}+\frac{\tau^a\theta^a-\beta}{2\pi
R}\right)$ for each $(c,b)$,\footnote{ It becomes more visible if we
diagonalize $\theta^aT^a$.  } and we can apply eq.~\eqref{eq_conv2}. We
show an example of the two-loop calculation in appendix
\ref{apx_floop_derv}.

There are four diagrams at the two-loop level\footnote{The one-loop
counter term for a wave function does not contribute to the Higgs
potential at the two-loop level since the $\theta$-dependence is
completely canceled between the propagator and the counter term.}. After
applying eq.~\eqref{eq_conv2}, we obtain the following expressions.

i) A fermion loop with a gauge boson ladder:
\begin{align}
 V^{2L}_{F,\rm eff}(\theta) =i
\begin{fmffile}{2L-F}
\begin{gathered}
\begin{fmfgraph}(50,50)
\fmfleft{i}
\fmfright{o}
\fmf{phantom,tension=5}{i,v1}
\fmf{phantom,tension=5}{v2,o}
\fmf{fermion,left,tension=0.4}{v1,v2,v1}
\fmf{photon}{v1,v2}
\fmfdot{v1,v2}
\end{fmfgraph} 
\end{gathered}
\end{fmffile}
&=6g^2\sum_{m_1,m_2}G_\ell(m_1,m_2)\int\frac{d^5p}{(2\pi)^5}
 \int\frac{d^5k}{(2\pi)^5}e^{-i2\pi R(p_5m_1+k_5m_2)}\nonumber\\
 &\hspace{10ex}\times \frac{(k+p)^Mk_M}{p^Np_N(k+p)^L(k+p)_Lk^Kk_K},\label{eq_floop}
\end{align}
where
\begin{equation}
 G_\ell(m_1,m_2)\equiv[e^{i\theta^cT^cm_1}]_{ba}
 \tr[e^{i(\theta^c\tau_\ell^c-\beta_\ell)m_2}\tau_\ell^a\tau_\ell^b].
\end{equation}

ii) A ghost loop with a gauge boson ladder:
\begin{align}
 V^{2L}_{c,\rm eff}(\theta) =i
\begin{fmffile}{2L-C}
\begin{gathered}
\begin{fmfgraph}(50,50)
\fmfleft{i}
\fmfright{o}
\fmf{phantom,tension=5}{i,v1}
\fmf{phantom,tension=5}{v2,o}
\fmf{ghost,left,tension=0.4}{v1,v2,v1}
\fmf{photon}{v1,v2}
\fmfdot{v1,v2}
\end{fmfgraph} 
\end{gathered}
\end{fmffile}
&=-\frac{1}{2}g^2\sum_{m_1,m_2}G_{\rm adj}(m_1,m_2)\int\frac{d^5p}{(2\pi)^5}
 \int\frac{d^5k}{(2\pi)^5}e^{-i2\pi R(p_5m_1+k_5m_2)}\nonumber\\
 &\hspace{10ex}\times\frac{(k+p)^Mk_M}{p^Np_N(k+p)^L(k+p)_Lk^Kk_K},
\end{align}
where
\begin{equation}
 G_{\rm adj}(m_1,m_2)\equiv[e^{i\theta^cT^cm_1}]_{ba}\tr[e^{i\theta^cT^cm_2}T^aT^b].
\end{equation}

iii) A gauge boson loop with a gauge boson ladder:
\begin{align}
 V^{2L}_{A1,\rm eff}(\theta) =i
\begin{fmffile}{2L-A1}
\begin{gathered}
\begin{fmfgraph}(50,50)
\fmfleft{i}
\fmfright{o}
\fmf{phantom,tension=5}{i,v1}
\fmf{phantom,tension=5}{v2,o}
\fmf{photon,left,tension=0.4}{v1,v2,v1}
\fmf{photon}{v1,v2}
\fmfdot{v1,v2}
\end{fmfgraph} 
\end{gathered}
\end{fmffile}
&=2g^2\sum_{m_1,m_2}G_{\rm adj}(m_1,m_2)\int\frac{d^5p}{(2\pi)^5}
 \int\frac{d^5k}{(2\pi)^5}e^{-i2\pi R(p_5m_1+k_5m_2)}\nonumber\\
 &\hspace{10ex}\times\frac{k^Mk_M+k^Mp_M+p^Mp_M}{p^Np_N(k+p)^L(k+p)_Lk^Kk_K}.
\end{align}

iv) Gauge boson loops connected by a four-point vertex:
\begin{align}
 V^{2L}_{A2,\rm eff}(\theta) =i
\begin{fmffile}{2L-A2}
\begin{gathered}
\begin{fmfgraph}(60,30)
\fmfleft{i}
\fmfright{o}
\fmf{phantom,tension=5}{i,i1}
\fmf{phantom,tension=5}{o,o1}
\fmf{photon,left,tension=0.4}{i1,v1,i1}
\fmf{photon,right,tension=0.4}{o1,v1,o1}
\fmfdot{v1}
\end{fmfgraph}
\end{gathered}
\end{fmffile}
&=-5g^2\sum_{m_1,m_2}G_{\rm adj}(m_1,m_2)\int\frac{d^5p}{(2\pi)^5}
 \int\frac{d^5k}{(2\pi)^5}e^{-i2\pi R(p_5m_1+k_5m_2)}\nonumber\\
 &\hspace{10ex}\times\frac{1}{k^Mk_Mp^Np_N}.
\end{align}

These loop integrals can be decomposed as
\begin{align}
 &\int\frac{d^5p}{(2\pi)^5}\int\frac{d^5k}{(2\pi)^5}e^{-i2\pi R(p_5m_1+k_5m_2)}
 \frac{ak^Mk_M+2bk^Mp_M+cp^Mp_M}{p^Np_N(k+p)^L(k+p)_Lk^Kk_K}\nonumber\\
 &\hspace{5ex}=-bF(m_1)F(m_2)-(a-b)F(m_1-m_2)F(m_2)-(c-b)F(m_1)F(m_2-m_1),
\end{align}
where
\begin{align}
 F(m)\equiv i\int\frac{d^5k}{(2\pi)^5}\frac{e^{-i2\pi Rk_5m}}{k^Kk_K}=
\begin{cases}
 \frac{1}{64\pi^5|m|^3R^3},&m\neq0,\\
 0,&m=0.
\end{cases}
\end{align}
These integrals are executed in appendix \ref{sec_integral}, where we
also show that $F(0)$ should vanish if we use a regularization that is
consistent with gauge invariance.

In appendix \ref{apx_proof_sym}, we see that $G_{\rm adj}(m_1,m_2)$
and $G_{\ell}(m_1,m_2)$ are symmetric under $m_1\leftrightarrow
m_2-m_1$, and that $G_{\rm adj}(m_1,m_2)$ is also symmetric under
$m_1\leftrightarrow-m_2$.

Using these, we get the two-loop effective potential as
\begin{align}
 V_{\rm eff}^{2L}(\theta)&=-3g^2\sum_\ell\sum_{m_1,m_2}G_\ell(m_1,m_2)
 [2F(m_1)F(m_2)-F(m_1)F(m_2-m_1)]\nonumber\\
 &\hspace{3ex}+\frac{9}{4}g^2\sum_{m_1,m_2}G_{\rm adj}(m_1,m_2)F(m_1)F(m_2).
\end{align}

As we can see, the result is finite.  The Abelian case can be obtained
by $T^a\to 0$ and $\tau^a_\ell\to Q_\ell$ with $Q_\ell$ being the
$U(1)$-charge of $\psi_\ell$. The result is consistent with the previous
works \cite{Maru:2006wa,Hosotani:2007kn}.

\section{Divergences at Higher-loop Level}\label{sec_higher}
In the previous section, we have seen that the Higgs effective potential
is finite up to the two-loop level.  At the one-loop level, the results
are finite because we need a non-zero winding number to get
$\theta$-dependent contributions. At the two-loop level, it is because
of the gauge invariance for the gauge boson self-energy. However, there
seems to be no reason that divergences should vanish at higher-loop
levels. Since the theory is non-renormalizable, we need infinite number
of counter terms, such as that for the four-Fermi operators. Connecting
the external lines of such counter terms, one can easily get
$\theta$-dependent contributions. Thus, if there is no non-trivial
cancellation, the Higgs effective potential depends on such counter
terms and hence on UV-theory. In this section, we show an example of
such divergences.

Since gauge theory in ${\bf M}^4\times S^1$ lies around the boundary of
renormalizable and non-renormalizable theories, the divergences appear
at rather higher-loop levels and it is a little hard to test the
finiteness explicitly. Thus, we increase the spacial dimension and
consider ${\bf M}^5\times S^1$. To improve visibility, we consider only
one massless Dirac fermion and suppress the flavor index, $\ell$. The
one-loop and the two-loop contributions are parallel to the previous
discussion and can be shown to be finite.

In this example, we concentrate on the four-Fermi operator and show that
the Higgs effective potential depends on its counter term.  The one-loop
corrections to the four-Fermi operator are log-divergent and the
divergent part is calculated as
\begin{align}
&\left.
\begin{fmffile}{FF1}
\begin{gathered}
\begin{fmfgraph}(60,30)
\fmfbottom{i1,d1,o1}
\fmftop{i2,d2,o2}
\fmf{fermion}{o1,v2,v1,i1}
\fmf{fermion}{o2,v4,v3,i2}
\fmf{photon,tension=0}{v1,v3}
\fmf{photon,tension=0}{v2,v4}
\end{fmfgraph}
\end{gathered}
\end{fmffile}
+
\begin{fmffile}{FF2}
\begin{gathered}
\begin{fmfgraph}(60,30)
\fmfbottom{i1,d1,o1}
\fmftop{i2,d2,o2}
\fmf{fermion}{o1,v2,v1,i1}
\fmf{fermion}{o2,v4,v3,i2}
\fmf{photon,tension=0}{v1,v4}
\fmf{photon,tension=0}{v2,v3}
\end{fmfgraph}
\end{gathered}
\end{fmffile}\right|_{\rm div}+({\rm crossed})\nonumber\\
&\hspace{3ex}=\frac{-ig^4}{768\pi^3}\frac{1}{\epsilon}
 \left[
 \gamma^L\gamma^N\gamma^M\tau^c\tau^a
 \right]_{\alpha\beta}
 \left[
 \gamma_M\gamma_N\gamma_L\tau^a\tau^c-\gamma_L\gamma_N\gamma_M\tau^c\tau^a
 \right]_{\gamma\delta}
 -(\alpha\leftrightarrow\gamma),
\end{align}
where $\alpha$ and $\gamma$ are spin indices of $\bar\psi$, and $\beta$
and $\delta$ are those of $\psi$. Here, $({\rm crossed})$ represents the
same diagrams with the fermion lines being crossed. We have used the
dimensional regularization and $\epsilon=3-D/2$ with $D$ being the
spacetime dimension.

To subtract the divergence, we need the following counter term;
\begin{equation}
 \mathcal L_{\rm CT}=\frac{\delta_{4F}}{2}
 \left[
 \bar\psi\gamma^M\gamma^N\gamma^L\tau^a\tau^b\psi
 \right]\left[
 \bar\psi\left(\gamma_M\gamma_N\gamma_L\tau^a\tau^b
 -\gamma_L\gamma_N\gamma_M\tau^b\tau^a\right)\psi
 \right],
\end{equation}
where
\begin{equation}
 \delta_{4F}=\frac{g^4}{768\pi^3}\frac{1}{\epsilon}+\delta_{4F}^{\rm fin}.
\end{equation}
Here, $\delta_{4F}^{\rm fin}$ represents finite renormalization and is
determined by UV-theory.\footnote{Since the four-Fermi operator is not
forbidden by any symmetry, $\delta_{4F}^{\rm fin}$ is arbitrary.}

By connecting the fermion lines of the counter term, we get a finite
contribution to the Higgs effective potential as
\begin{align}
 V_{\rm CT}(\theta)&=\sum_{m_1\neq0}\sum_{m_2\neq0}
 \frac{\delta_{4F}^{\rm fin}\mathcal N}{8\pi^{16} R^{10}m_1^5m_2^5}\nonumber\\
 &\hspace{5ex}\times\left\{2\tr\left[\tau^ae^{i(\theta^b\tau^b-\beta)m_1}\right]
 \tr\left[\tau^ae^{i(\theta^b\tau^b-\beta)m_2}\right]
 +\tr\left[\tau^ae^{i(\theta^b\tau^b-\beta)m_1}
 \tau^ae^{i(\theta^b\tau^b-\beta)m_2}\right]\right\}.
\end{align}
It is non-vanishing and has non-trivial $\theta$-dependence.  For
example, in the ${\rm SU}(2)$ gauge theory with a fermion in the fundamental
representation, the above contribution can be expressed as
\begin{equation}
V_{\rm CT}^{\mathcal N=2}(\theta)=\sum_{m_1\neq0}\sum_{m_2\neq0}
 \frac{3\delta_{4F}^{\rm fin}e^{-i\beta(m_1+m_2)}}{8\pi^{16} R^{10}m_1^5m_2^5}
 \cos\left(\frac{m_1+m_2}{2}\sqrt{(\theta^1-2\beta)^2+(\theta^2)^2+(\theta^3)^2}\right),
\end{equation}
which is indeed non-vanishing.  Notice that there is only one Wilson
line phase in the ${\rm SU}(2)$ case, which corresponds to $\sqrt{\dots}$ in
the above equation. Thus, the Higgs effective potential inevitably
depends on UV-theory. It falsifies the conjecture of all-order
finiteness in this model.\footnote{ This result is not strong enough to
rule out the all-order finiteness for an Abelian case since $V_{\rm
CT}(\theta)$ vanishes identically. In addition, we do not exclude the
possibility that the effective potential becomes finite at all-order
after the inclusion of $\delta_{4F}$. However, it is beyond our criteria
of finiteness.}

Notice that the above contribution cannot be canceled by any other
diagrams. To see this, let us integrate out the fermions since we are
not interested in external fermion lines. Since all of the fermion legs
should be connected, the contribution of the counter term to any
operator in the effective action starts at the three-loop level. There
is only one diagram that contributes to the Higgs potential at the
three-loop level, which is what we calculated above. One might afraid
that the renormalization of the gauge coupling constant or the self
energies may affect the result. However, since the modification of the
renormalization starts at the three-loop level, their contribution to
the Higgs potential appears at four- or higher-loop levels.

The above example implies that there is no special mechanism that
prevents the Higgs effective potential to diverge. Since there are
infinite number of counter terms, we expect that the effective potential
is generically divergent at three- or higher-loop levels also in other
models.

Although the Higgs effective potential seems to be divergent, it is
notable that the divergence is suppressed at least at the three-loop
level. Since the gauge theory is non-renormalizable, we expect that it
is UV-completed at a scale that is not so far from $1/R$. Thus, such a
higher-loop suppression can be strong enough to explain the little
hierarchy between these scales.

\section{Summary}\label{sec_summary}
In this paper, we have investigated the finiteness of the Higgs
effective potential in a non-Abelian GHU model defined on ${\bf
M}^4\times S^1$. Although the model is non-renormalizable, the Higgs
effective potential is known to be finite at the one-loop level and it
has been conjectured that it might be free from divergences at all orders
in perturbative expansions. However, the calculation of the effective
potential beyond the one-loop level has been a technical challenge and
only the two-loop calculation in an Abelian model is available in the
literature \cite{Maru:2006wa,Hosotani:2007kn}.

To overcome the technical difficulties, we presented a powerful method
to calculate the loop integrals in the GHU, {\it compactification by
superposition}. We express a loop integral and sum in ${\bf M}^4\times
S^1$ as a superposition of loop integrals in ${\bf M}^5$, which allows
us to remove all the matrix valued objects from the integrals. The Higgs
dependence of the potential is then expressed as phase factors in
association with the superposition, where the periodicity of the Higgs
potential is manifest.

Using the method, we have determined the effective potential up to the
two-loop level in the non-Abelian model, which turned out to be finite.

We have also discussed that the Higgs effective potential are
generically divergent at the three- or higher-loop levels. As an
example, we have considered an ${\rm SU}(\mathcal N)$ gauge theory on ${\bf
M}^5\times S^1$. We have seen that the one-loop correction to the
four-Fermi operator is divergent and we need a counter term to
renormalize the theory. Then, we have explicitly shown that the Higgs
effective potential depends on the counter term at the three-loop level,
which falsifies the conjecture of the all-order finiteness for this
model. It seems that this feature is generic since there are
infinite number of counter terms and one can easily generate the Higgs
potential by connecting their legs.

Although the effective potential seems to be divergent, it is found to
be suppressed at least at the three-loop level. Such higher-loop
suppression is still useful to explain the hierarchy between the scale
of the GHU and that of a UV cutoff.

\begin{acknowledgements}
This work was supported by Grant-in-Aid for Scientific research from
the Ministry of Education, Science, Sports, and Culture (MEXT), Japan,
No. 16H06492 [J.H. and Y.S.]. The work of J.H. is also supported by
World Premier International Research Center Initiative (WPI
Initiative), MEXT, Japan.
\end{acknowledgements}

\appendix
\section{Proof of Identities}
\subsection{Proof of eq.~\eqref{eq_conv2}}\label{apx_proof_main}
Let $\Theta$ be a Hermitian matrix. Then, for an arbitrary analytic
function, $S(\dots)$, the following identity holds.
\begin{equation}
 \frac{1}{2\pi R}\sum_{n=-\infty}^{\infty}S
 \left(\frac{n}{R}+\frac{\Theta}{2\pi R}\right)
 =\sum_{n=-\infty}^{\infty}e^{i\Theta n}
 \int_{-\infty}^{\infty} \frac{dk_5}{2\pi}S(k_5)e^{-i2\pi Rk_5n}.
\end{equation}

\underline{\large Proof}

We first diagonalize $\Theta$ as
\begin{equation}
 U^{-1}\Theta U={\rm diag}~(v_1,v_2,\cdots),
\end{equation}
with unitary matrix $U$.  Since $S(\dots)$ is an analytic function, we have
\begin{equation}
 \left[S\left(\frac{n}{R}+\frac{\Theta}{2\pi R}\right)\right]_{ab}
 =U_{ac}S\left(\frac{n}{R}+\frac{v_c}{2\pi R}\right)U_{cb}^{-1}.
\end{equation}
Inserting an identity, we get
\begin{equation}
 \left[S\left(\frac{n}{R}+\frac{\Theta}{2\pi R}\right)\right]_{ab}
 =\int_{-\infty}^{\infty} \frac{dk_5}{2\pi}S(k_5)U_{ac}2\pi\delta
 \left(k_5-\frac{n}{R}-\frac{v_c}{2\pi R}\right)U_{cb}^{-1}.
\end{equation}
Since we have
\begin{equation}
 \sum_{n=-\infty}^{\infty}2\pi\delta\left(p-\frac{n}{R}\right)
 =2\pi R\sum_{n=-\infty}^{\infty}e^{-i2\pi Rpn},
\end{equation}
we get
\begin{equation}
 \sum_{n=-\infty}^{\infty}\left[S\left(\frac{n}{R}+\frac{\Theta}{2\pi R}\right)\right]_{ab}
 =2\pi R\int_{-\infty}^{\infty} \frac{dk_5}{2\pi}S(k_5)
 \sum_{n=-\infty}^{\infty}U_{ac}e^{-i2\pi R
 \left(k_5-\frac{v_c}{2\pi R}\right)n}U_{cb}^{-1}.
\end{equation}
Using
\begin{equation}
 e^{-i2\pi R\left(k_5-\frac{v_c}{2\pi R}\right)n}\delta_{cd}
 =\left[e^{-i2\pi R\left(k_5-\frac{{\rm diag}~(v_1,v_2,\cdots)}{2\pi R}\right)n}\right]_{cd},
\end{equation}
we get
\begin{equation}
 \sum_{n=-\infty}^{\infty}S\left(\frac{n}{R}+\frac{\Theta}{2\pi R}\right)
 =2\pi R\int_{-\infty}^{\infty} \frac{dk_5}{2\pi}
 \sum_{n=-\infty}^{\infty}e^{-i2\pi R\left(k_5-\frac{\Theta}{2\pi R}\right)n}S(k_5).
\end{equation}
\subsection{Proof of eq.~\eqref{eq_move}}\label{apx_proof_com}
 Let $\tau^a$'s be an arbitrary representation of ${\rm SU}(\mathcal N)$, $\lambda^a$'s
 be constants and $S(\dots)$ be an arbitrary analytic function. Then, the following identity holds;
\begin{equation}
 S(\lambda^a\tau^a)\tau^b=\tau^c\left[S\left(\lambda^a\tau^a+\lambda^aT^a\right)\right]_{cb},
\end{equation}
where the indices in the subscript are those for $T^a$, not for $\tau^a$.

\underline{\large Proof}

Since $S(\dots)$ can be expanded locally, it is enough to prove for the
case where $S(\dots)$ is a monomial function.  Since we have
\begin{equation}
 [\lambda^a\tau^a,\tau^b]=\tau^c(\lambda^aT^a_{cb}),
\end{equation}
we have
\begin{align}
 (\lambda^a\tau^a)^n\tau^b
 &=(\lambda^a\tau^a)^{n-1}\tau^c[\delta^{cb}\lambda^a\tau^a+\lambda^aT^a_{cb}]\nonumber\\
 &=\cdots=\tau^c[\lambda^a\tau^a+\lambda^aT^a]_{cb}^n.
\end{align}
Since the above holds for each term of the Taylor series, the same holds for $S(\dots)$.
\subsection{Symmetries of $G_{\rm adj}$ and $G_\ell$}\label{apx_proof_sym}
 Let $\tau^a$'s be an arbitrary representation of ${\rm SU}(\mathcal N)$ and $\lambda^a$'s
 and $\bar\lambda^a$'s be constants. Then, the following identities hold;
 \begin{align}
  \left[e^{i\lambda^cT^c}\right]_{ba}\tr\left[e^{i\bar\lambda^c\tau^c}\tau^a\tau^b\right]
 &=\left[e^{i(\bar\lambda^c-\lambda^c)T^c}\right]_{ba}\tr\left[e^{i\bar\lambda^c\tau^c}\tau^a\tau^b\right],\\
  \left[e^{i\lambda^cT^c}\right]_{ba}\tr\left[e^{i\bar\lambda^cT^c}T^aT^b\right]
 &=\left[e^{-i\bar\lambda^cT^c}\right]_{ba}\tr\left[e^{-i\lambda^cT^c}T^aT^b\right].
 \end{align}

\underline{\large Proof}

The first identity can be shown by using the identity of appendix
\ref{apx_proof_com}. We have
 \begin{align}
  \left[e^{i\lambda^cT^c}\right]_{ba}\tr\left[e^{i\bar\lambda^c\tau^c}\tau^a\tau^b\right]
 &=\left[e^{i\lambda^cT^c}\right]_{ba}
 \left[e^{i\bar\lambda^cT^c}\right]_{da}\tr\left[\tau^de^{i\bar\lambda^c\tau^c}\tau^b\right]\nonumber\\
 &=\left[e^{i\lambda^cT^c}\right]_{ba}\left[e^{-i\bar\lambda^cT^c}\right]_{ad}
 \tr\left[\tau^de^{i\bar\lambda^c\tau^c}\tau^b\right]\nonumber\\
 &=\left[e^{i(\bar\lambda^c-\lambda^c)T^c}\right]_{db}
 \tr\left[e^{i\bar\lambda^c\tau^c}\tau^b\tau^d\right].
 \end{align}

The second identity can be shown as
\begin{align}
 \left[e^{i\lambda^cT^c}\right]_{ba}\tr\left[e^{i\bar\lambda^cT^c}T^aT^b\right]
 &=\left[e^{i\lambda^cT^c}\right]_{ba}\left[e^{i\bar\lambda^cT^c}\right]_{cd}T^a_{de}T^b_{ec}\nonumber\\
 &=\left[e^{-i\lambda^cT^c}\right]_{ab}\left[e^{-i\bar\lambda^cT^c}\right]_{dc}T^c_{be}T^d_{ea}\nonumber\\
 &=\left[e^{-i\bar\lambda^cT^c}\right]_{dc}\tr\left[e^{-i\lambda^cT^c}T^cT^d\right].
\end{align}

\section{Momentum Integrals}\label{sec_integral}
\subsection{Momentum Integrals with a Spacial Shift Operator}
In this appendix, we calculate
\begin{equation}
 \mathcal I=\int \frac{d^Dk}{(2\pi)^D}(-k^Mk_M+2p^Mk_M+m^2-i\epsilon)^{-s}e^{-i2k^M x_M}.
\end{equation}

From the definition of the gamma function, we have
\begin{equation}
 W^{-s}=\frac{i^s}{\Gamma(s)}\int^\infty_0dt~e^{-iWt}t^{s-1},
\end{equation}
for $\Im(W)<0$ and $\Re(s)>0$.
Using this, we have
\begin{align}
 \mathcal I&=\frac{i^s}{\Gamma(s)}\int^\infty_0dt~t^{s-1}
 \int \frac{d^Dk}{(2\pi)^D}e^{i(k^Mk_M-2p^Mk_M-m^2+i\epsilon)t-i2k^M x_M}\nonumber\\
 &=\frac{i^s}{\Gamma(s)}e^{-i2p^Mx_M}\int^\infty_0dt~t^{s-1}e^{-i\frac{x^Mx_M}{t}-i(p^Mp_M+m^2)t-\epsilon t}
 \int \frac{d^Dk}{(2\pi)^D}e^{ik^Mk_Mt}\nonumber\\
 &=\frac{i^{s-D/2+1}}{\Gamma(s)(4\pi)^{D/2}}e^{-i2p^Mx_M}
 \int^\infty_0dt~t^{s-D/2-1}e^{-i\frac{x^Mx_M}{t}-i(p^Mp_M+m^2)t-\epsilon t}.
\end{align}

The integral can be evaluated as
\begin{align}
 \lim_{\delta\to+0}\int^\infty_0dt~t^{r-1}e^{-iBt+i\frac{C}{t}-\epsilon t-\frac{\delta}{t}}
 &=2(-i)^{r/2}\frac{C^{r/2}}{[i(B-i\epsilon)]^{r/2}}K_r\left(-2i^{3/2}\sqrt{iC(B-i\epsilon)}\right),
\end{align}
for $\epsilon>0,~C>0$, where $K_n(z)$ is the modified Bessel function of
the second kind. Here, we introduced a regulator $\delta>0$. 

When $0<\Re(s)$ and $p^Mp_M+m^2\neq0$, the integral is convergent and is evaluated as
\begin{align}
 \mathcal I &= \frac{2i^{s/2-D/4+1}}{(4\pi)^{D/2}\Gamma(s)}
 \frac{e^{-i2p^Mx_M}(-x^Mx_M)^{s/2-D/4}}{[i(p^Mp_M+m^2-i\epsilon)]^{s/2-D/4}} \nonumber\\
 &\hspace{3ex}\times K_{s-D/2}
 \left(-2i^{3/2}\sqrt{i(p^Mp_M+m^2-i\epsilon)(-x^Mx_M)}\right).
\end{align}
Notice that
\begin{equation}
 K_{n+1/2}(x)=K_{-n-1/2}(x)=\left(\frac{\pi}{2x}\right)^{1/2}e^{-x}
 \sum_{r=0}^n\frac{(n+r)!}{r!(n-r)!}(2x)^{-r},
\end{equation}
with $n$ being a positive integer.

When $p^Mp_M+m^2=0$, we need to take $B\to0$ before $\epsilon\to0$, which gives
\begin{equation}
 \lim_{B\to0}\lim_{\delta\to+0}\int^\infty_0dt~t^{r-1}e^{-iBt+i\frac{C}{t}-\epsilon t-\frac{\delta}{t}}
 =C^r(-i)^r\Gamma(-r)+\mathcal O(\epsilon).
\end{equation}

When $0<\Re(s)<\frac{D}{2}$ and $p^Mp_M+m^2=0$, it becomes
\begin{equation}
  \mathcal I = \frac{i}{(4\pi)^{D/2}}
 \frac{\Gamma\left(\frac{D}{2}-s\right)}{\Gamma(s)}
 e^{-i2p^Mx_M}(-x^Mx_M)^{s-D/2}.
\end{equation}
\subsection{Proof of $F(0)=0$}
We assume a regularization that has the following features.
\begin{itemize}
 \item All the integrals become finite.
 \item Invariance under the shifts of loop momenta.
 \item Independence of the signs of loop momenta.
 \item Gauge invariance, $p_M\Pi^{MN}(p)=0$.
\end{itemize}
Then, the following identity holds;
\begin{equation}
 F(0)\equiv i\int\frac{d^5k}{(2\pi)^5}\frac{1}{k^Kk_K}=0.
\end{equation}

\underline{\large Proof}

Let us define
\begin{align}
 \Lambda^3&\equiv -iF(0)=\int\frac{d^5k}{(2\pi)^5}\frac{1}{k^Mk_M},\\
 \Xi(p)&\equiv\int\frac{d^5k}{(2\pi)^5}\frac{1}{(k+p/2)^M(k+p/2)_M(k-p/2)^N(k-p/2)_N}.
\end{align}
Then, we have the following relations;
\begin{align}
  &\int\frac{d^5p}{(2\pi)^5}\Xi(p)=(\Lambda^3)^2,
 \label{xilambda}\\
 &\int\frac{d^5k}{(2\pi)^5}\frac{k^Mk^N}{(k+p/2)^L(k+p/2)_L(k-p/2)^K(k-p/2)_K}\nonumber\\
 &\hspace{5ex}=\left(\frac{1+x}{5}\eta^{MN}-x\frac{p^Mp^N}{p^Lp_L}\right)
 \left[\Lambda^3-\frac{p^Lp_L}{4}\Xi(p)\right],
\end{align}
where $x$ is a constant, which will be determined later.

At the one-loop level, the divergent corrections to the gauge boson
self-energy are given by
\begin{align}
 &\left.
\begin{minipage}{5.5em}
\includegraphics[width=5.5em]{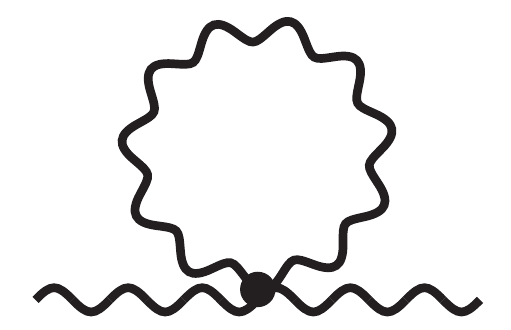}
\end{minipage}+
\begin{fmffile}{SE-G}
\begin{gathered}
\begin{fmfgraph}(60,40)
\fmfleft{i}
\fmfright{o}
\fmf{photon,tension=4}{i,v1}
\fmf{photon,tension=4}{v2,o}
\fmf{photon,left,tension=1}{v1,v2,v1}
\fmfdot{v1,v2}
\end{fmfgraph}
\end{gathered}
\end{fmffile}+
\begin{fmffile}{SE-C}
\begin{gathered}
\begin{fmfgraph}(60,40)
\fmfleft{i}
\fmfright{o}
\fmf{photon,tension=4}{i,v1}
\fmf{photon,tension=4}{v2,o}
\fmf{ghost,left,tension=1}{v1,v2,v1}
\fmfdot{v1,v2}
\end{fmfgraph}
\end{gathered}
\end{fmffile}
\right|_{\rm div}\nonumber\\
&\hspace{3ex}=\frac{g^2}{2}\Xi(p)\tr \left[T^aT^d\right]
 \left[-3\left(\frac{1+x}{5}p^Lp_L\eta^{MN}-xp^Mp^N\right)+4\left(p^Lp_L\eta^{MN}-p^Mp^N\right)\right]_{db}\nonumber\\
 &\hspace{6ex}+3g^2\Lambda^3
 \tr \left[T^aT^d\right]\left[2\left(\frac{1+x}{5}\eta^{MN}-x\frac{p^Mp^N}{p^Lp_L}\right)-\eta^{MN}\right]_{db},\nonumber\\
 &\left.
\begin{fmffile}{SE-F}
\begin{gathered}
\begin{fmfgraph}(60,40)
\fmfleft{i}
\fmfright{o}
\fmf{photon,tension=4}{i,v1}
\fmf{photon,tension=4}{v2,o}
\fmf{fermion,left,tension=1}{v1,v2,v1}
\fmfdot{v1,v2}
\end{fmfgraph}
\end{gathered}
\end{fmffile}
\right|_{\rm div}\nonumber\\
 &\hspace{3ex}=-2g^2\Xi(p)\tr \left[\tau_\ell^a\tau_\ell^d\right]
 \left[-\left(\frac{1+x}{5}p^Lp_L\eta^{MN}-xp^Mp^N\right)+p^Lp_L\eta^{MN}-p^Mp^N\right]_{db}\nonumber\\
 &\hspace{6ex}-4g^2\Lambda^3\tr \left[\tau_\ell^a\tau_\ell^d\right]
 \left[2\left(\frac{1+x}{5}\eta^{MN}-x\frac{p^Mp^N}{p^Lp_L}\right)-\eta^{MN}\right]_{db},
\end{align}
where the external lines have indices of $(M,a)$ and $(N,b)$ and
$p_M=\left(p_\mu,p_5+\frac{\theta^aT^a}{2\pi R}\right)$ is the external
momentum.

The gauge invariance requires
\begin{equation}
 p^2\Xi(p)(4x-1)-2(3+8x)\Lambda^3=0.
\end{equation}
Its possible solutions, which are also consistent with eq.~\eqref{xilambda},
are
\begin{align}
 &\Lambda^3=0,~x=\frac{1}{4},\\
 &\Xi(p)=\frac{\Lambda^3}{p^Mp_M},~x=-\frac{7}{12},\\
 &\Xi(p)=0,~\Lambda^3=0. 
\end{align}
The second one does not regularize the integral for $p=0$ and thus is
not suitable for regularization. The last one is a special case of the
first one.

Thus, we conclude
\begin{align}
 &\Lambda^3=0,~x=\frac{1}{4}.
\end{align}
Notice that, if we use the dimensional regularization, $\Lambda^3,x$ and
$\Xi(p)$ are explicitly calculated as
\begin{equation}
 \Lambda^3=0,~x=\frac{1}{4},~\Xi(p)=-\frac{i}{128\pi}\sqrt{-p^Mp_M}.
\end{equation}

\section{Example of the Two-loop Calculation}\label{apx_floop_derv}
This appendix is dedicated to deduce eq.~\eqref{eq_floop}, the contribution
from the two-loop diagram with a fermion loop;
\begin{equation}
V^{2L}_{F,\rm eff}(\theta) =i
\begin{fmffile}{2L-F}
\begin{gathered}
\begin{fmfgraph}(50,50)
\fmfleft{i}
\fmfright{o}
\fmf{phantom,tension=5}{i,v1}
\fmf{phantom,tension=5}{v2,o}
\fmf{fermion,left,tension=0.4}{v1,v2,v1}
\fmf{photon}{v1,v2}
\fmfdot{v1,v2}
\end{fmfgraph} 
\end{gathered}
\end{fmffile}.
\end{equation}

First, we apply the Feynman rules to the diagram and get
\begin{multline}
V^{2L}_{F,\rm eff}(\theta)
=\frac{i}{2}\frac{1}{2\pi R}\sum_{n_1}\int\frac{d^4p}{(2\pi)^4}
\frac{1}{2\pi R}\sum_{n_2}\int\frac{d^4k}{(2\pi)^4}
\left[\frac{-i\eta_{MN}}{p^2-(\frac{n_1}{R}
+\frac{\theta^cT^c}{2\pi R})^2}\right]_{ab}\\
\times(-1)\tr\left[\frac{i}{\cancel p+\cancel k-\gamma_5(\frac{n_1+n_2}{R}
+\frac{\theta^c\tau_\ell^c-\beta_\ell}{2\pi R})}ig\gamma^M\tau_\ell^a
\right.\\
\left.\times\frac{i}{\cancel k-\gamma_5(\frac{n_2}{R}+\frac{\theta^c\tau_\ell^c-\beta_\ell}{2\pi R})}
ig\gamma^N\tau_\ell^b\right],
\end{multline}
where the trace operates on both $\gamma^M$'s and $\tau_\ell^a$'s.

Next, we use eq.~\eqref{eq_move} to rearrange the integrand into an
analytic function of $(\frac{n_1}{R}+\frac{\theta^cT^c}{2\pi R})$ and
$(\frac{n_2}{R}+\frac{\theta^c\tau_\ell^c-\beta_\ell}{2\pi R})$. After
$\tau_\ell^a$ being shifted to the left, $\tr[\dots]$ in the above
equation becomes
\begin{equation}
g^2\tr\left[\tau_\ell^e\left(\frac{1}{\cancel p+\cancel k-\gamma_5(\frac{n_1+n_2}{R}
+\frac{\theta^c\tau_\ell^c-\beta_\ell+\theta^cT^c}{2\pi R})}\right)_{ea}\gamma^M
\frac{1}{\cancel k-\gamma_5(\frac{n_2}{R}+\frac{\theta^c\tau_\ell^c-\beta_\ell}{2\pi R})}
\gamma^N\tau_\ell^b\right],
\end{equation}
where the subscript, $ea$, shows the indices for $T^a$.

Now, we can apply eq.~\eqref{eq_conv2} to
$(\frac{n_1}{R}+\frac{\theta^cT^c}{2\pi R})$ and
$(\frac{n_2}{R}+\frac{\theta^c\tau_\ell^c-\beta_\ell}{2\pi R})$. Notice that
we can treat these variables independently since they can be
diagonalized simultaneously. We obtain
\begin{multline}
V^{2L}_{F,\rm eff}(\theta)=\frac{i}{2}\sum_{m_1}\int\frac{d^5p}{(2\pi)^5}
\sum_{m_2}\int\frac{d^5k}{(2\pi)^5}e^{-i2\pi R(p_5m_1+k_5m_2)}
\left[e^{i\theta^cT^cm_1}\right]_{ab}\frac{-i\eta_{MN}}{p^Lp_L}\\
\times(-g^2)\tr\left[e^{i(\theta^c\tau_\ell^c-\beta_\ell)m_2}\frac{1}{\gamma^J(p+k)_J}\gamma^M
\frac{1}{\gamma^I k_I}\gamma^N\tau_\ell^b\tau_\ell^a
\right].
\end{multline}
Working out the trace of the gamma matrices, we obtain
\begin{multline}
 V^{2L}_{F,\rm eff}(\theta)=6g^2\sum_{m_1,m_2}\left[e^{i\theta^cT^cm_1}\right]_{ab}
\tr\left[e^{i(\theta^c\tau_\ell^c-\beta_\ell)m_2}\tau_\ell^b\tau_\ell^a\right]\hspace{20ex}\\
 \times\int\frac{d^5p}{(2\pi)^5} \int\frac{d^5k}{(2\pi)^5}e^{-i2\pi R(p_5m_1+k_5m_2)}
\frac{(p+k)^Mk_M}{p^Lp_L(p+k)^J(p+k)_Jk^Ik_I}.
\end{multline}
\bibliographystyle{apsrev4-1}
\bibliography{ghu}
\end{document}